\documentclass[aip,apl,12pt]{revtex4-1}
\usepackage{graphicx}

\begin{document}

\title{On-chip spectroscopy with thermally-tuned high-Q photonic crystal cavities}

\author{Andreas C. Liapis}
\email[Electronic address: ]{andreas.liapis@gmail.com}

\author{Boshen Gao}
\author{Mahmudur R. Siddiqui}
\affiliation{The Institute of Optics, University of Rochester, Rochester, NY 14627, USA.}

\author{Zhimin Shi}
\affiliation{Department of Physics, University of South Florida, Tampa, FL 33620, USA.}

\author{Robert W. Boyd}
\affiliation{The Institute of Optics, University of Rochester, Rochester, NY 14627, USA.}
\affiliation{Department of Physics and School of Electrical Engineering and Computer Science, University of Ottawa, Ottawa, ON K1N 6N5, Canada.}

\date{\today}

\begin{abstract}
Spectroscopic methods are a sensitive way to determine the chemical composition of potentially hazardous materials. Here, we demonstrate that thermally-tuned high-Q photonic crystal cavities can be used as a compact high-resolution on-chip spectrometer. We have used such a chip-scale spectrometer to measure the absorption spectra of both acetylene and hydrogen cyanide in the 1550 nm spectral band, and show that we can discriminate between the two chemical species even though the two materials have spectral features in the same spectral region. Our results pave the way for the development of chip-size chemical sensors that can detect toxic substances.
\end{abstract}


\maketitle 

\begin{figure}
	\includegraphics[width=3in]{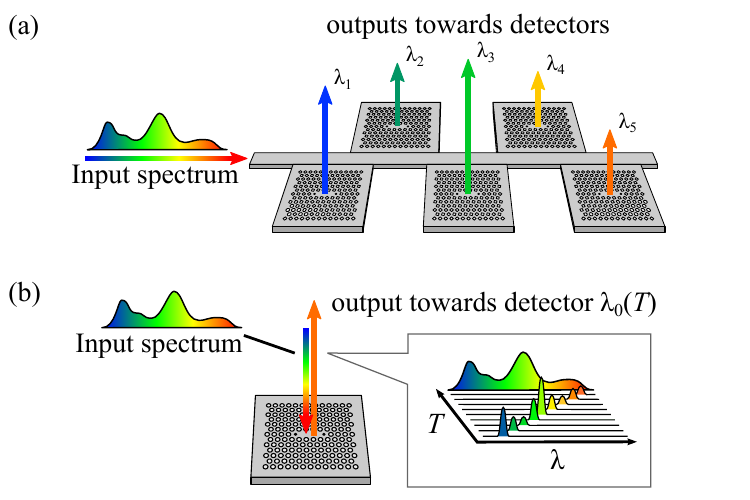}
	\caption{Operating principle of a PhC cavity spectrometer using (a) an array of sequentially tuned cavities or (b) a single dynamically-tunable cavity.\label{Operating_Principle}}
\end{figure}

The past decade has seen a marked increase in the demand for compact, integrated devices that can reliably identify chemical and biological agents. Miniaturized spectrometers, in particular, have received much attention.
Broadly speaking, two categories of on-chip spectrometers exist: Firstly, there are dispersive spectrometers, such as arrayed-waveguide grating (AWG) spectrometers,\cite{AWG,Zhimin_AWG} superprism spectrometers,\cite{GaTech,Boshen_OpEx} grating-based spectrometers,\cite{Grating,Lipson} and on-chip implementations of the Mach-Zehnder geometry.\cite{MZ} However, these devices tend to be rather large, with a resolution in wavenumbers (cm$^{-1}$) typically given by $1/L$, where $L$ is the characteristic size of the dispersive region in cm.
Spectrometers can also be constructed by cascading optical elements with narrow-band spectral responses, such as Fabry-P\'erot filters,\cite{Wang_OL} grating couplers,\cite{Pervez_OpEx} microdonut resonators,\cite{Xia_OpEx} photonic crystal cavities\cite{Deotare_IEEE,Gan_APL,Meng_APL} etc. The operating principle of such a spectrometer is shown schematically in Fig.~\ref{Operating_Principle}(a), and can be described as follows:

The spectrum to be measured is injected into a bus waveguide that is coupled to a number of resonators, each tuned to a different wavelength. The light scattered by each resonator, either vertically, as in Gan {\it et al.},\cite{Gan_APL} or into output waveguides, as in Xia {\it et al.},\cite{Xia_OpEx} is monitored and used to reconstruct the input spectrum. The resolution of such a spectrometer is given by the spectral width of the frequency response of the individual elements, which is typically of the order of 10 GHz. Conversely, the total number of resolvable spectral lines, that is, the number of resonators that can be used, is limited by their free spectral range, that is, the separation between adjacent resonance frequencies of the cavity. Note that, in practice, the total number of resonators is limited by the loss introduced by each element for the non-resonant frequency component.
The drawback of cascaded spectrometers is that they require additional optics to couple light into and out of the resonators, which increases the device size and complicates its operation.

In this letter, we demonstrate high-resolution spectroscopy based on a single tunable high-Q silicon photonic crystal (PhC) cavity that both emits into and is excited from the vertical direction, thus eliminating the need for a bus waveguide entirely [Fig.~\ref{Operating_Principle}(b)]. In addition, we demonstrate that such a cavity can readily be used to discriminate between the absorption spectra of different chemical species.
Two gases are considered here, acetylene ($^{12}$C$_2$H$_2$) and hydrogen cyanide (H$^{12}$CN). These are commonly used as wavelength references, with strong absorption lines in the vicinity of 1530 nm. Both gases are used industrially, but hydrogen cyanide is highly toxic.
Our results therefore pave the way for the development of chip-size chemical sensors that can detect the presence of toxic substances.

\begin{figure*}
	\includegraphics{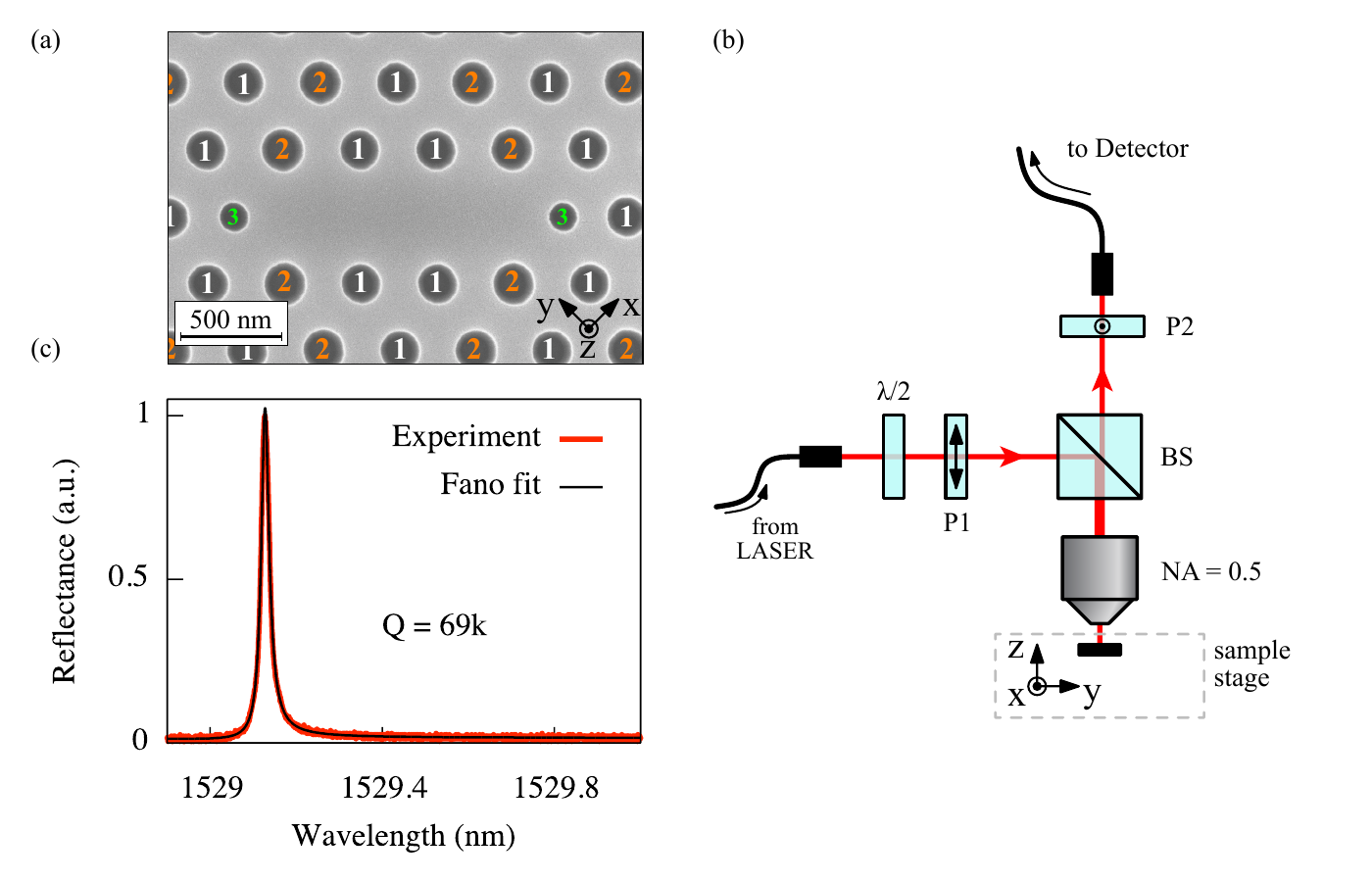}
	\caption{(a) Scanning-electron micrograph of an L-3 PhC cavity optimized for vertical out-coupling. The holes are labeled according to their design radius. In this example, $r_1 = 89$ nm, $r_2 = 93$ nm, $r_3 = 62$ nm. Additionally, the holes labeled `3' are shifted outwards by 62 nm. (b) Schematic of the setup used to characterize PhC cavities via resonant scattering. P1 and P2: crossed polarizers, $\lambda/2$: half-wave plate, BS: beam splitter. Note that the light incident on the PhC cavity is  y-polarized, and that only light that is x-polarized is being collected by the detector. (c) Measured resonance of a typical L-3 PhC cavity, displaying a Q of $\sim$ 69000.~\label{Setup}}
\end{figure*}

We have fabricated PhC cavities with enhanced vertical coupling using a design protocol similar to that of Tran {\it et al.}\cite{Tran_PRB} These cavities consist of a triangular lattice of air holes on a thin silicon membrane, with three holes omitted, forming an L-3 cavity. The cavity's terminating holes, labeled `3' in Fig.~\ref{Setup}(a), are reduced in size and slightly displaced to increase the cavity Q-factor. Additionally, the radii of selected holes surrounding the cavity, labeled `2' in Fig.~\ref{Setup}(a), are slightly increased to modify the cavity's far-field emission pattern and thus facilitate coupling in the out-of-plane direction. These PhC cavities were fabricated on commercial silicon-on-insulator wafers (SOITEC). The patterns were defined by electron-beam lithography on a positive-tone resist (ZEP 520), and transferred to the Si membrane by inductively-coupled plasma etching. After removal of the remaining resist, the buried oxide layer was undercut by wet chemical etching. A scanning-electron micrograph of one such cavity is shown in Fig.~\ref{Setup}(a).

The fabricated cavities are characterized using resonant-scattering spectroscopy.\cite{McCutcheon_APL} In our experimental setup [Fig.~\ref{Setup}(b)], the cavity axis is oriented at a $45^\circ$ angle with respect to the polarization of the incident light, which is set by a half-wave plate and a polarizer. A second polarizer, oriented at a $90^\circ$ angle with respect to the incident polarization, controls the polarization of light that reaches the detector. As a result of this arrangement, stray light reflected from the sample surface is suppressed, and only light that has scattered resonantly through the cavity is collected.
The resonantly-scattered intensity can then be fit with a Fano lineshape, from which we extract the resonance wavelength $\lambda_0$ and quality factor Q of the cavity.\cite{Galli_APL} For a typical cavity, Q is approximately 69000 [Fig.~\ref{Setup}(c)].

\begin{figure}
	\includegraphics[width=3in]{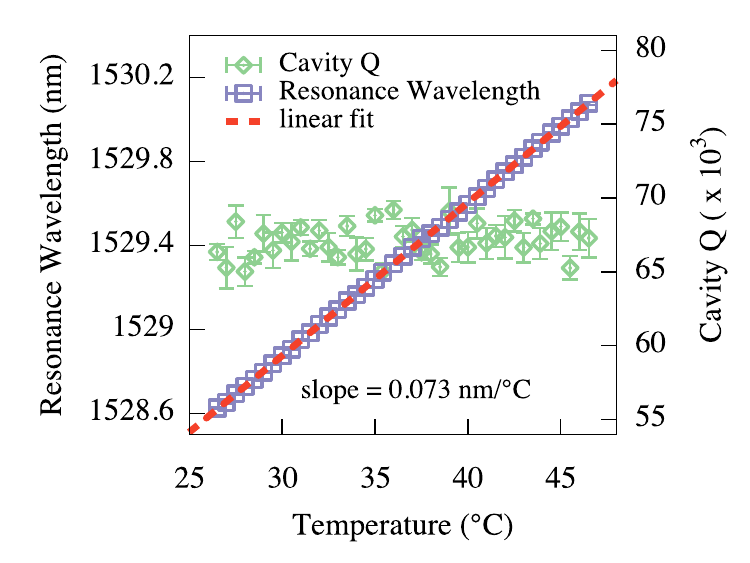}%
	\caption{Resonance wavelength and cavity Q as functions of temperature for an optimized L3 PhC cavity. A tuning range of 1.4 nm is demonstrated.\label{Thermal}}%
\end{figure}

To examine the dependence of the resonance wavelength on temperature, a heating element is attached to the sample mounting block. The temperature is monitored with a thermocouple and is stabilized using a proportional-integral-derivative feedback circuit.
At each temperature point we record the resonance wavelength $\lambda_0$ and coupling efficiency $\eta$, which is given by the ratio of scattered to incident optical power. A linear dependence of $\lambda_0$ on temperature is observed, with a slope of 0.07~nm$/^\circ$C (Fig.~\ref{Thermal}), which is consistent with what one would expect based on the temperature dependence of the refractive index of silicon.\cite{wild2004temperature} The useful tuning range is approximately 1.5 nm for a temperature range of 20~$^\circ$C. The cavity Q-factor remains constant within 5\% over this tuning range. 

\begin{figure*}
\includegraphics[width=6in]{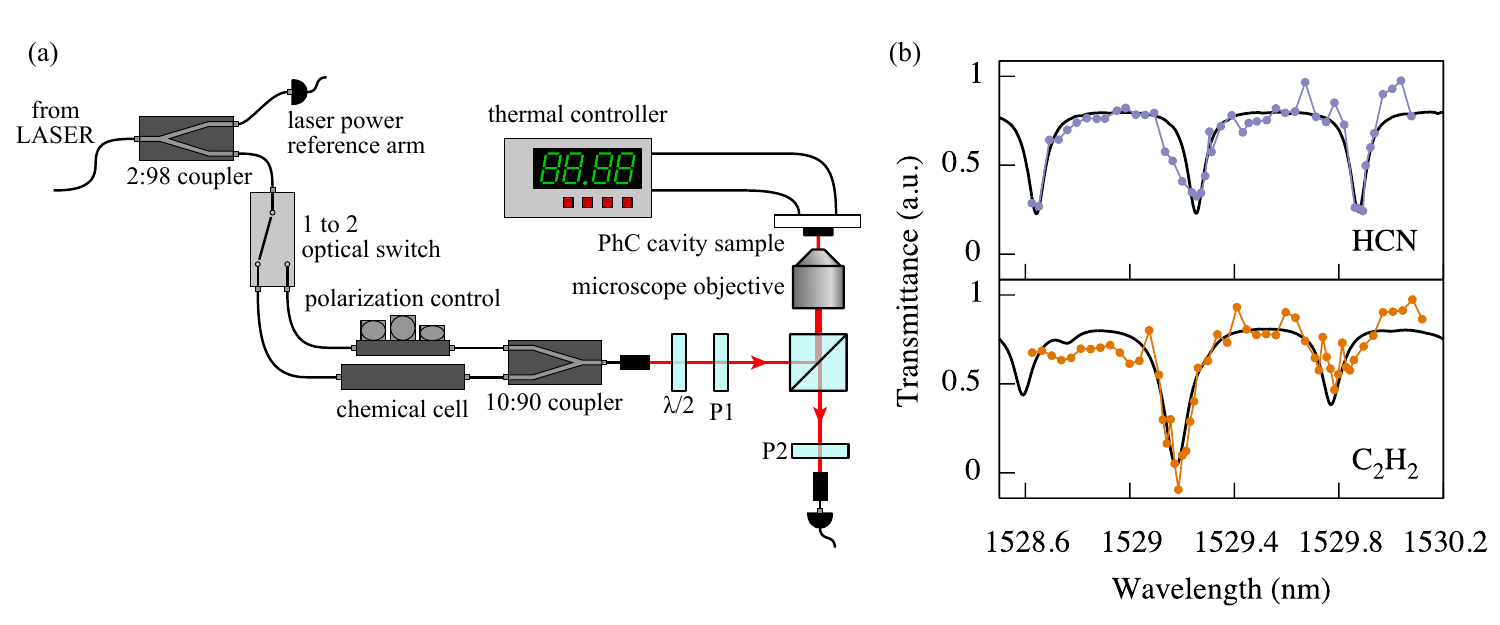}
\caption{(a) Schematic of the setup used for spectroscopic measurements. (b) Measured transmission spectra of the hydrogen cyanide (top) and acetylene (bottom) gas cells. The transmission of the two gas cells, measured using a tunable laser, is also shown (solid lines). \label{spect}}
\end{figure*}

Once characterized, the cavity can be used to perform spectroscopy with a spectral resolution of 0.02~nm (or, equivalently, 3~GHz or 0.1~wavenumbers at a wavelength range near 1550 nm), which is comparable to the resolution of a tabletop spectrometer. The cavity is illuminated with the absorption spectrum we wish to measure, and the scattered optical power $O$ is recorded as the temperature of the chip is tuned. To facilitate integration into our setup [Fig.~\ref{spect}(a)], we use commercial fiber-coupled gas cells (Wavelength References Inc). The acetylene cell is 5.5-cm long and filled to a pressure of 740 Torr, while the HCN cell is 16.5-cm long and filled to 150 Torr.
Fig.~\ref{spect}(b) shows the transmission of the two gas cells, obtained by scanning a tunable laser over the wavelength range 1528.5 -- 1530.2 nm, which covers lines P6 -- P8 of acetylene and R5 -- R7 of hydrogen cyanide.
The spectra measured using the PhC cavity, which are given by the ratio $O/\eta$, are also shown.  
Note that the alternating strength of the absorption lines seen in the spectrum of acetylene is due to the triple degeneracy of the antisymmetric rotational levels of this linear symmetric molecule, which is a result of the nuclear moments $I_H = 1/2$ and $I_C = 0$.\cite{Herzberg_1945} Hydrogen cyanide, which is linear but not symmetric, does not display such an intensity alternation.

We see that the resolution of our cavity spectrometer is adequate to resolve neighboring absorption lines of acetylene and hydrogen cyanide.
In principle, even better performance could be achieved by carefully optimizing the design of the PhC cavities to further increase their quality factor.
Similar cavities were recently demonstrated with Q factors exceeding one million.\cite{Lai_APL} 
Note, however, that in our current setup, the spectral resolution is limited not only by the cavity Q, but also by the stability of the thermal controller.

It is worth noting that by using a bus waveguide and multiple resonators, one could in principle perform parallel monitoring to decrease the data acquisition time.  However, to construct a spectrometer of this type, we would need to fabricate a series of cavities whose resonances are equally spaced and span the target spectral range. The quality factor and resonance wavelength of PhC cavities can be chosen by appropriately tuning the design parameters, but not with arbitrary precision. Control is ultimately limited by the inherent imperfections of the fabrication process. Unavoidable imperfections in the resulting structure, such as sidewall roughness and lithographic inaccuracies, lead to variations in resonance frequency even among nominally identical cavities.\cite{Portalupi_PRB} These variations become more pronounced for high-Q cavities, for which the uncertainty in resonance wavelength can be much larger than their resonance linewidth. As a result, unless these variations are corrected for, fabrication disorder limits the practical resolution of the spectrometer, as one is confined to using low-Q cavities only. Even though multiple techniques have been demonstrated, \cite{hennessy2005tuning,faraon2008local,hennessy2006tuning,lee2009local,chen2011selective} controllable post-fabrication tuning of individual PhC cavities remains a technical challenge.

In summary, we have demonstrated an implementation of a chip-scale spectrometer based on a dynamically-tuned photonic crystal cavity. By utilizing a cavity optimized for vertical out-coupling, we avoid having to use a bus waveguide to couple light into and out of the cavity, and thus drastically decrease the footprint of the device. By monitoring the light reflected by this cavity as a function of temperature we can extract spectral information with a resolution of 0.02 nm. Our device therefore overcomes the $1/L$ limit for the spectral resolution of dispersive spectrometers. A standard grating spectrometer would need a footprint with a scale size of at least 10 cm to achieve comparable resolution (0.1 cm$^{-1}$).

\begin{acknowledgments}
The authors would like to thank Antonio Badolato, Sebastian A. Schulz, David D. Smith, and Jerry Kuper for many fruitful discussions, and acknowledge the assistance of Emily Conant in the early stages of the experiment. 
This work was supported by the US Defense Threat Reduction Agency--Joint Science and Technology Office for Chemical and Biological Defense (Grant No. HDTRA1-10-1-0025), by the National Aeronautics and Space Administration (under contract  NNX15CM47P), and by the Canada Excellence Research Chairs Program.
Fabrication of the photonic crystal nanocavities was performed at the Cornell NanoScale Facility, a member of the National Nanotechnology Coordinated Infrastructure (NNCI), which is supported by the National Science Foundation (Grant ECCS-15420819).
\end{acknowledgments}
~\\
\bibliography{biblio.bib}

\end{document}